\documentclass[twocolumn,showpacs,aps,pre]{revtex4}
\usepackage{epsfig}
\usepackage{amssymb}

\begin{document}
\title{Analytic study of clustering in shaken granular material using
zero-range processes}
\author{J\'anos T\"or\"ok}
\affiliation{Institute of Physics, University Duisburg-Essen,
D-47048 Duisburg, Germany}
\affiliation{Department of Chemical Information Technology, Budapest
University of Technology and Economics, H-1111 Budapest, Hungary}

\begin{abstract}
We show that models used to described granular clustering due to
vertical shaking belong to the class of zero-range processes. This
correspondence allows us to derive exactly in a very easy and
straightforward manner a number of properties of the models like
particle distribution functions, phase diagram, and characteristic
time of clusterization.
\end{abstract}


\maketitle

\section{Introduction}

Non-equilibrium phase transitions were observed in many simple systems
\cite{Schutzrev,Zia,Privman}. Recently, it was also found that shaken
granular material exhibits clustering depending on the shaking
strength \cite{Eggers}. In his paper Jens Eggers suggested a model for
the description of the clustering of vertically shaken granular
material. Originally, it was introduced for the two box setup. Later
it and its modified version were used to describe experiments having
more boxes \cite{Weele,Meer1,Meer2,Droz1,Droz2,Shim1,Shim2}.

The analytic studies of the above mentioned papers were difficult and
in many cases did not give general results. In the case of exclusion
models, it was proved \cite{Erel} that the correspondence to an
already solved model, namely the zero-range process \cite{Shim2} can
be of great help, as many results can be obtained directly. 

The aim of this paper to derive analytically for the {\em general}
case the probability distribution, phase diagram and the coarsening
time using the above correspondence. We note that specific cases with
small number of boxes are already solved
\cite{Weele,Meer1,Meer2,Droz1,Droz2,Shim1,Shim2}.

\section{The model}

The experimental setup looks as follows. The system consists of a
container separated into boxes by walls which are open upwards. The
whole system is then vertically shaken and particles can hop above the
walls from one compartment to the other. The model of Eggers
\cite{Eggers} defines particle fluxes between the boxes based upon the
physical properties of the system like shaking strength and local
particle density, etc.. Two different steady states were found, both
in the experiments and in the model: a homogeneous where the
containers hold roughly equal number of particles, and a condensed
steady state where one compartment contains nearly all particles.

We first define the zero-range process following \cite{Evans}: We
consider a one-dimensional finite lattice of $L$ sites and periodic
boundary conditions. The total particle number is denoted by $N$.

The dynamics of the system is given by rates $u(n)$, at which a
particle leaves a site. The hopping rates $u(n)$ depend {\em only} on
the number of particles on the site of departure and external
parameters but independent of the properties of the target site. Here,
we consider only the symmetric case when the hopping to the left or to
the right is equally probable.

The important attribute of the zero-range process is that it yields a
steady state described by a product measure. By this it is meant that
the steady state probability $P(\{n_\mu\})$ of finding the system in
configuration $\{n_1,n_2,\dots n_L\}$ is given by a product of factors
$f(n_\mu)$ that are called marginals
\begin{equation}
P(\{n_\mu\})={1\over Z(L,N)}\prod_{\mu=1}^L f(n_\mu),
\end{equation}
where $Z(L,N)$ is the normalization factor. For the zero-range process
$f(n_\mu)$ is given by
\begin{equation}
f(n)=\cases{\displaystyle \prod_{m=1}^n{1\over u(m)} & for $n\geq 1$ \cr
1 & for $n=0$ }.
\end{equation}
The marginals are defined up to a multiplicative factor.

It is important to note that the steady state particle distribution
can be formally read off generally for all $N$ and $L$. This will
allow us to calculate the general phase diagrams.

\begin{figure}
\epsfig{file=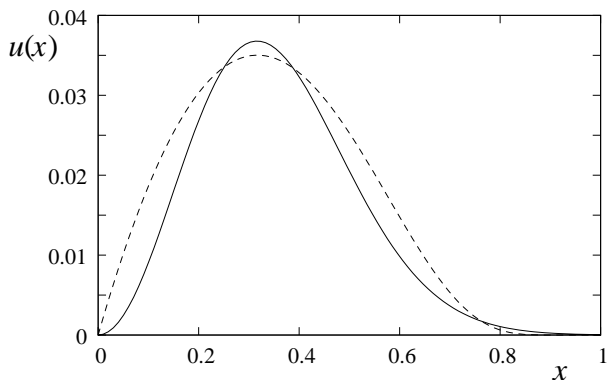,width=8cm}
\caption{\label{Fig:un} 
Sample hopping probabilities. Solid line {\em E} dashed {\em LD}
model.
}
\end{figure}

The Eggers model \cite{Eggers} has a quadratic temperature dependence
while Lipowski and Droz \cite{Droz1} uses a linear form. For all
quantities we use an index {\em E} for Eggers model and {\em LD} for
the Lipowski-Droz model.  The fluxes are defined in the following way
\begin{eqnarray}\label{Eq:un}
u_E(n)&=&An^2e^{-Bn^2}\cr
u_{LD}(n)&=&{n\over N}\exp\left(-1/[T_0+\Delta(1-n/N)]\right),
\end{eqnarray}
where $A$, $B$, $T_0$, and $\Delta$ contain physical quantities. It is
of importance that $B$ is proportional to the inverse shaking strength
while $\Delta$ increases with the strength of the agitation. The $u(n)$
functions are compared on Fig. \ref{Fig:un}.

Our main purpose in this paper is to study the clusterization
process of this model. Furthermore, as cases with $L=2,3$
\cite{Weele,Meer1,Meer2,Droz1,Droz2,Shim1,Shim2} are already solved we
focus on the general case with at least moderate $L$ and large $N\gg1$
particle number.

\section{Particle distribution}

Before calculating the general particle distribution function we
remark that in both models only two different steady states may be
stable: (i) The symmetric, where all sites contain the same amount of
particles, and (ii) The asymmetric, or condensed steady state, where
one site contains more particles than the others which have equal
number of particles.

The reason why only those two are stable is that in the steady state
all sites must have the same flux. The hopping probability curve has a
maximum thus sites may be on both sides of the maximum at the
positions with equal flux. However, when the flux has negative
derivate only one site may be at a given position because more would
be unstable against fluctuations.

Thus we calculate only the probability distribution function for the
case where $L-1$ sites contain the same number of particles $a$ the
remaining site thus holds $b=N-(L-1)a$ particles. The marginals take
the form
\begin{eqnarray}
f_E(n)&=&{\exp
\left({B\over 6}n(n+1)(2n+1)\right)\over A(n!)^2}\cr
f_{LD}(n)&=&{N!\over n!}\left[T_0+\Delta\left(1-{n\over
N}\right)\right]^{-N/\Delta}.
\end{eqnarray}
We are lucky that for both models the unnormalized probability
distribution can be obtained in a closed form:
\begin{eqnarray}
P_E(\{a,\dots a,b\})&=&{
\exp[B({L-1\over 3}a^3+{L-1\over 2}a^2+{L-1\over 6}a)]
\over A^La!^{2(L-1)}b!^2} \cr
&&\times \exp[B(b^3/3+b^2/2+b/6)]\cr
&&\cr
P_{LD}(\{a,\dots a,b\})&=&{N!^L\over a!^{L-1}b!}
\left[T_0+\Delta\left(1-{a\over N}\right)\right]^{-N(L-1)\over\Delta}\cr
&&\times\left[T_0+\Delta\left(1-{b\over N}\right)\right]^{-N/\Delta}.
\label{Eq:Pgen}
\end{eqnarray}

\section{Phase diagram}

We already discussed that both models have two possible stable steady
states which give rise to four different cases on the phase diagram:
I) only the symmetric state is stable, II) only the asymmetric state
is stable, III) both states are stable but asymmetric is more
probable, IV) both states are stable but symmetric is more probable.
In this section we determine the three lines separating these states.

On line III-IV the symmetric solution loses its stability. In the
symmetric steady state all sites contain $N/L$ particles. Only a flux
curve with positive derivate may be stable for many sites thus $N/L$
must be less than the position of the maximum of the flux curve. This
gives the following two relations
\begin{eqnarray}\label{Eq:symstab}
(E) && B=L^2/N^2\cr
(LD) && T_0=\sqrt{\Delta/L}-\Delta(1-1/L).
\end{eqnarray}

The other two line can only be determined through the probability
distribution function.

\subsection{The {\em LD} model}

Let us define the particle ratios: $x=a/N$, $\beta=b/N$. The
following equation describes the extremal points of the probability
distribution function
\begin{equation}\label{Eq:nmeq}
\ln{\beta\over x}={1\over T_0+\Delta(1-\beta)}-{1\over
T_0+\Delta(1-x)}.
\end{equation}
It is obvious to see that $\beta=x$ is always a solution and that the
above equation is $N$ independent. The exact position of the
clusterized state can only be determined numerically.

An important question is whether the transition from symmetric to
asymmetric steady state is continuous or not. The third order Taylor
expansion of Eq. (\ref{Eq:nmeq}) around the symmetric solution of
$x=1/L$ shows that if $L>2$ no continuous transition is possible as
there is no solution of the equation. While for $L=2$ continuous
solution is possible if $\Delta<2/3$. This result was also shown with
a general argument in \cite{Shim1}. It is important to reiterate
because in the following calculations are valid for $x\ll1$ which can
only be used in case of discontinuous transitions.

The line I-IV, the limit of the stability of the asymmetric steady
state can be derived from Eq. (\ref{Eq:nmeq}). Assuming that $x\ll1$
and the fact that at the point where other roots except from $\beta=x$
appear the value and the derivate of both sides of the equation should
be equal. This results in the following implicit equation
\begin{equation}
0=\ln\left({\Delta(L-1)\over T_0^2}+{\Delta\over (T_0+\Delta)^2}\right)
+{T_0^2+\Delta(T_0-1)\over T_0(T_0+\Delta)}.
\end{equation}

The line III-IV, can not be exactly determined because we do not know
exactly the roots of Eq. (\ref{Eq:nmeq}). Because of the previous
arguments a good estimate can however, be found by assuming $x=0$ and
$\beta=1$ which is a very good approximation in most parts of the
phase space. The relation $P_{LD}(0,\dots0,N)=P_{LD}(N/L,\dots N/L)$
gives an implicit equation for $\Delta$ and $T_0$
\begin{equation}
0={(T_0+\Delta(L-1)/L)^L\over T_0(T_0+\Delta)^{(L-1)}}-L^\Delta.
\end{equation}

\begin{figure}
\epsfig{file=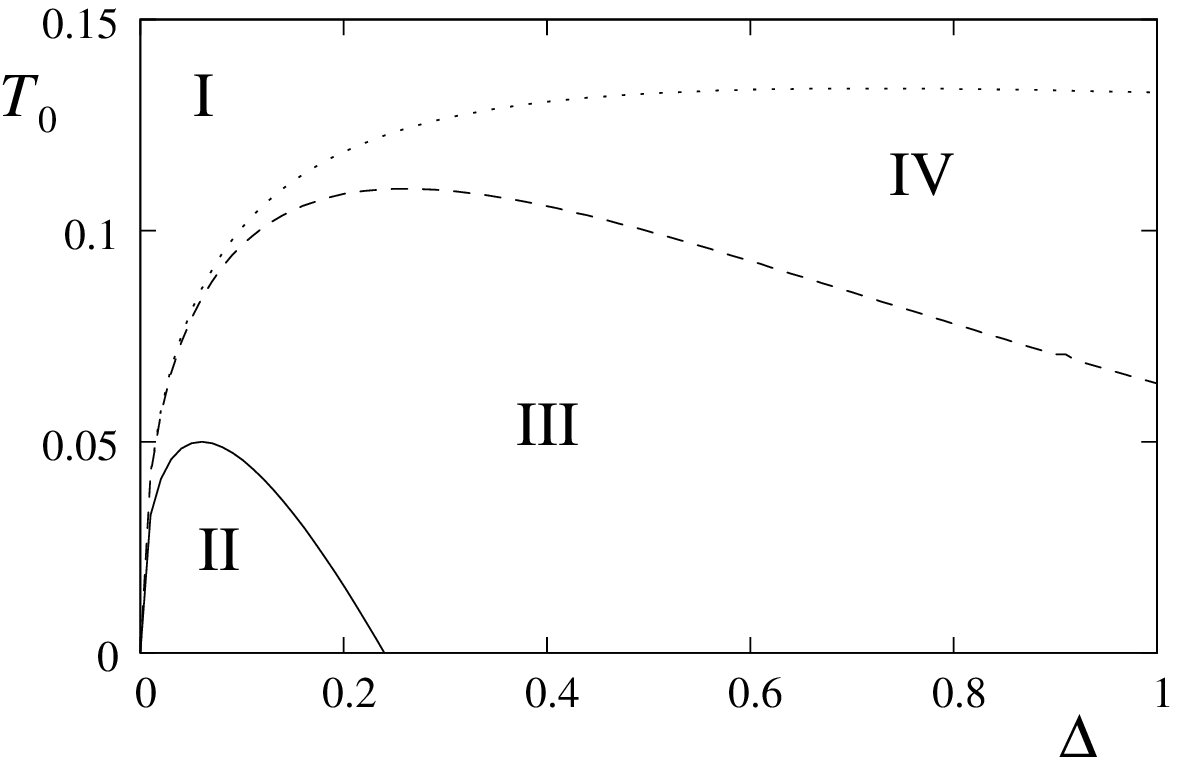,width=8cm}
\epsfig{file=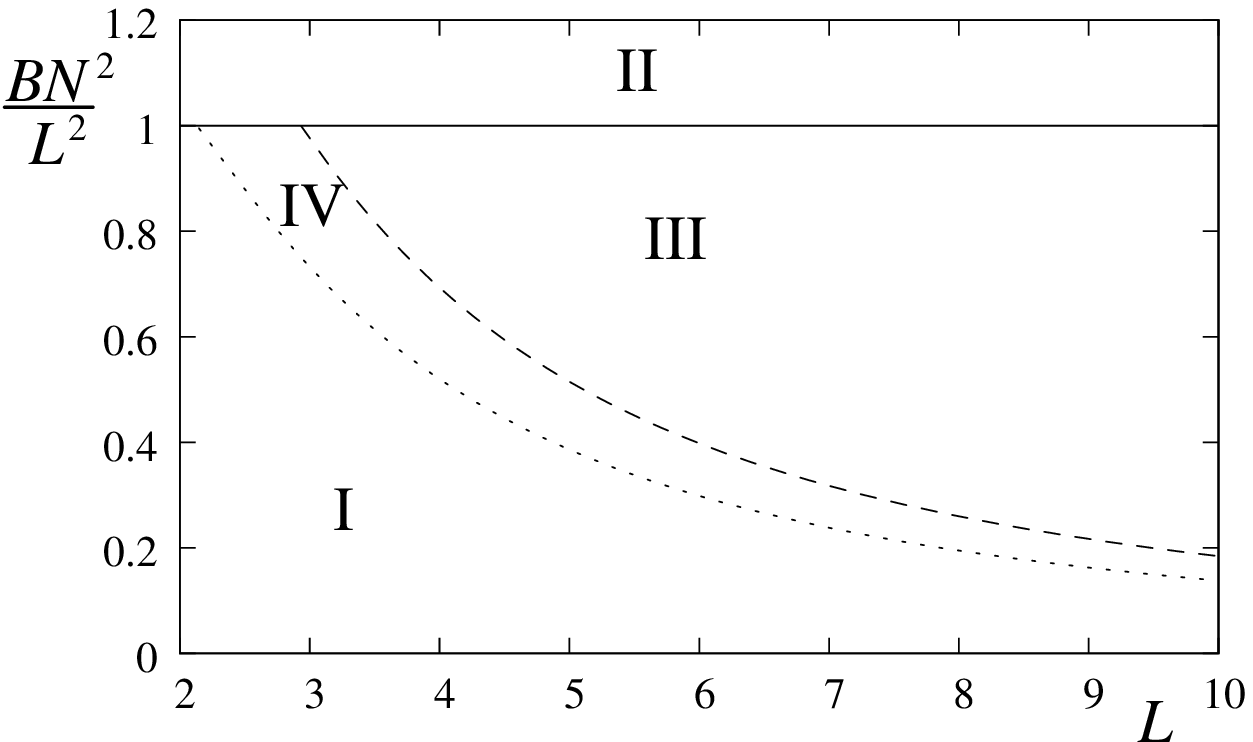,width=8cm}
\caption{\label{Fig:phase} 
Phase diagrams. Top (a) {\em LD} for $L=6$ bottom (b) model {\em E}.
}
\end{figure}

An example phase diagram for $L=6$ can be seen on Fig.
\ref{Fig:phase} a.

\subsection{The {\em E} model}

Following the same route as for the {\em LD} model we derive the
equation for the position of the steady states in the leading order of
$N$.
\begin{equation}
(\beta^2-x^2)B N^2=2\ln(\beta/x)+{\cal O}(1/N).
\end{equation}
The $N$ dependence of the equation disappears if one chooses $B\propto
N^{-2}$. It is also obvious that the above equation has the solution
of $x=\beta$. Other solutions must satisfy $x<[2(L-1)]^{-1}$. This
does not allow a continuous phase transition for $L>2$ but continuous
transition always exists for $L=2$ as was shown in \cite{Weele}.

Line I-IV can be determined from the above results
\begin{equation}
BN^2=8\ln L+{\cal O}(1).
\end{equation}

Line III-IV can be determined from the equality of the probability
distribution functions as
\begin{equation}
BN^2=6\ln L +{\cal O}(1).
\end{equation}

The above two equations suffer from strong corrections to the leading
order $\ln L$ term. However, this accuracy is enough the get a clear
picture of the phase diagram which is shown on Fig. \ref{Fig:phase}
b.

\section{Coarsening time}

The coarsening process in general drives the system to have bigger but
a smaller number of macroscopic clusters until only one prevails. As
the clusters become bigger the coarsening process slows down
enormously because the size of the clusters to be dissolved is bigger
and the relative flux is much smaller. Thus, in general it is enough
to study the competition of the last two big clusters \cite{Schutz} to
get the asymptotic behaviour of the coarsening time $\tau$.

The above picture applies only if the if the homogeneous steady state
is not stable. In the opposite case, where the symmetric steady state
is also stable a fluctuation large enough is needed to drive the
system out of the potential well for which the time needed is in
general proportional to $\exp(-N^2)$. Here we study only the case when
the system is in section II of the phase space.

For obvious reasons, the two big clusters are generally at $\sim L/2$
distance. If a big cluster looses a particle to its neighbour the
particle starts a random walk. If the distance between the two
clusters is $L/2$ then the probability of reaching the other is $2/L$
and it takes $L^2/4$ timesteps. Thus the coarsening time can be
written as
\begin{equation}
\tau= {L^3\over 8}\int_1^{N/2-x_l}{dn\over u(N/2-n)-u(N/2+n)}
\end{equation}

The coarsening time is be dominated by the state when both clusters
have almost the same number of particles. Thus, we use the first order
Taylor series around this point:
\begin{eqnarray}
\tau_{LD}&\simeq& L^3{\exp(2/(2T_0+\Delta))(2T_0+\Delta)^2\over
2\Delta-(2T_0+\Delta)^2}N\ln(N/2)\cr
\tau_{E}&\simeq&{L^3\exp(BN^2/4)\over AN^3B}\ln(N/2)
\end{eqnarray}

\begin{figure}
\epsfig{file=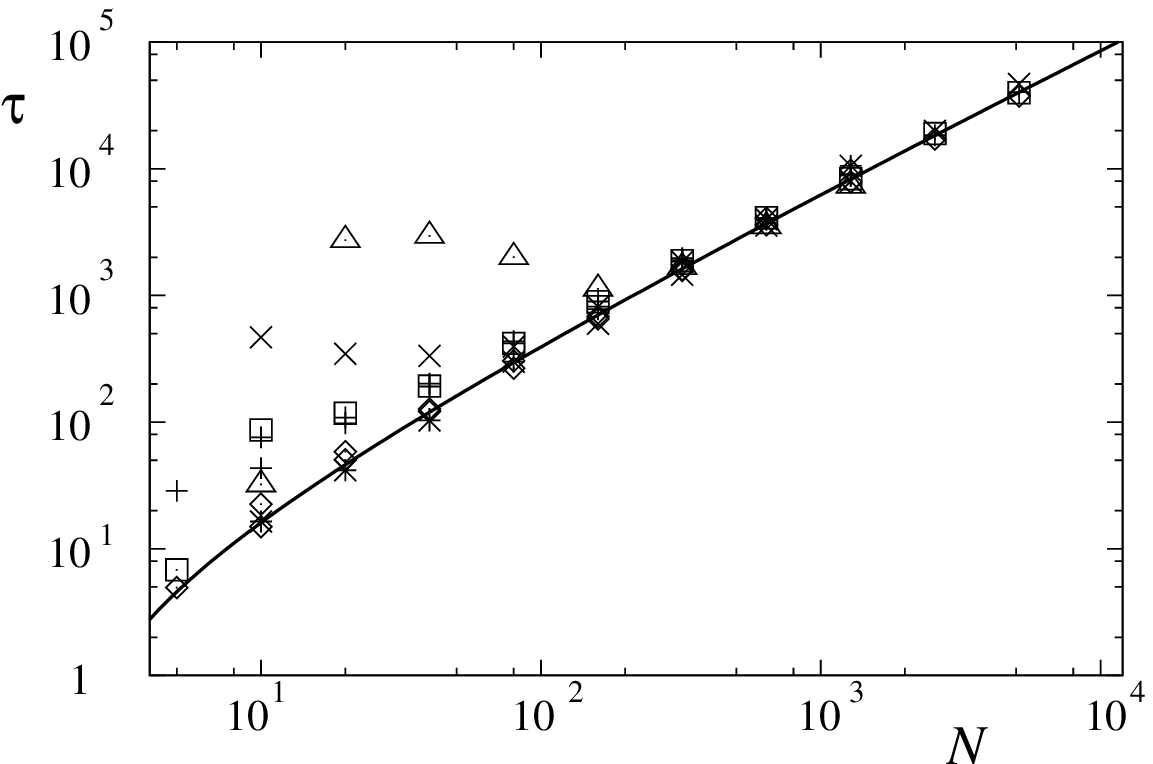,width=8cm}
\epsfig{file=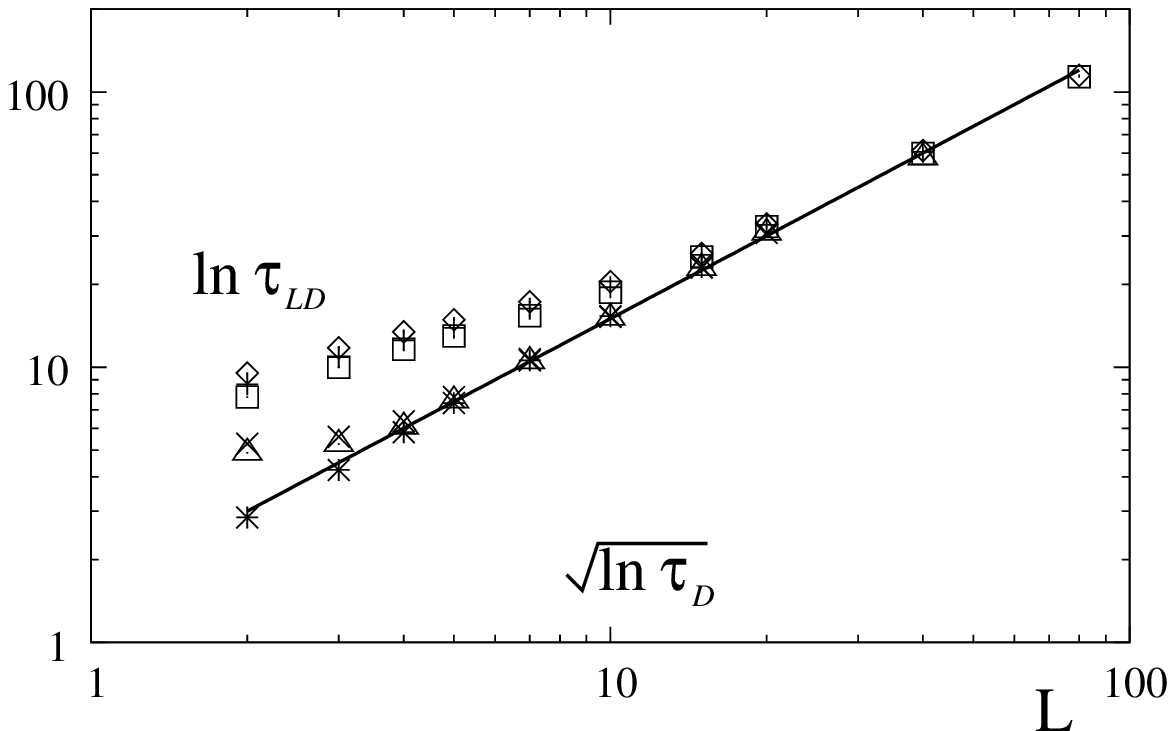,width=8cm}
\caption{\label{Fig:tL} 
(a) Top: Coarsening time vs. particle number. Box number
$L=2,3,5,10,20$ for both models scaled together. The solid line is
$N\ln(N/2)$. The stronger deviations for small particle numbers are
for the {\em E} model.
\hfill\break
(b) Bottom: Coarsening time vs. box number. System sizes
$N=80,160,320$ are scaled together for both models. Parameters were
chosen as $B\propto L^2$, $T_0\propto d\propto 1/L$, The solid line
is a linear. Curves with higher deviations for small box numbers are
for the {\em LD} model and $\ln\tau$ is plotted. For
the {\em E} model $\sqrt{\ln\tau_D}$ is shown.
}
\end{figure}

For the {\em E} model we have to identify the $N$ dependence of $A$.
This parameter sets the speed of the hopping but in all previous
calculations it was unimportant. We show that in the transition regime
it should scale as $1/N^2$ exactly as $B$. One reason could be the
physical derivation of Eggers \cite{Eggers}, where one can look up the
correspondence of $A$ and $B$, and find that they have the same $N$
dependence. The other reason is that we want to have the same order of
fluxes in different systems. The value of the maximum of $u(n)$ is
$AN^2$ so it is obvious to have $A\propto1/N^2$ in the transition
regime. Finally, we can get the same $N$ scaling as for the {\em LD}
model namely
\begin{equation}
\tau\propto N\ln(N/2),
\end{equation}
which is shown on Fig. \ref{Fig:tL}.

The $L$ dependence of the coarsening time can also be read off.
For the {\em E} model in region II of the phase space $B$ has a $L$
dependence faster than $L^2$. It is easy to see that if $B\propto
L^\gamma$ then the coarsening time grows as
$\tau_{E}\propto\exp(L^\gamma)$. For the {\em LD} model the dominating
term will also be the exponential as in region II of the phase space
both $\Delta$ and $T_0$ should decrease at least as $1/L$ for which
$\tau_{LD}\propto\exp(L)$. The results are tested against numerical
simulations on Fig. \ref{Fig:tL} b.

As we already mentioned at the model definition $B$ is inversely
proportional to the shaking strength, while $\Delta$ is proportional
to the agitation. From physical point of view the $L$ scaling of the
coarsening time is also the same for both models in spite of all
differences.

\section{Conclusion}

In this paper we showed that models proposed to described
clustering of shaken granular material belongs to the group of
zero-range processes which is very efficient in describing
non-equilibrium phase transitions.

The probability distribution which describes the steady state of the
system can be read off in a straightforward manner. This allows us to
determine the phase diagram analytically for both models in the
general case, for $L,N\ll1$.

The coarsening time is also calculated for both models and tested
against numerical simulations. The two models have qualitatively
different phase diagram but surprisingly similar coarsening time
scaling. The particle number scaling was found to be the same linear,
the box number dependence was found to be exponential.

Part of this work was supported by OTKA F047259 and by the
German-Hungarian Cooperation Fund.

\end{document}